# Isotropic Pauli-Limited Superconductivity in the Infinite-Layer Nickelate Nd$_{0.775}$Sr$_{0.225}$NiO$_2$

Bai Yang Wang,*[1,2] Danfeng Li,[2,3] Berit H. Goodge,[4] Kyuho Lee,[1,2] Motoki Osada[2,5], Shannon P. Harvey[2,3], Lena F. Kourkoutis,[4,6] Malcolm R. Beasley[3], and Harold Y. Hwang*[2,3]

[1]Department of Physics, Stanford University, Stanford, CA 94305, United States.

[2]Stanford Institute for Materials and Energy Sciences, SLAC National Accelerator Laboratory, Menlo Park, CA 94025, United States.

[3]Department of Applied Physics, Stanford University, Stanford, CA 94305, United States.

[4]School of Applied and Engineering Physics, Cornell University, Ithaca, NY 14853, United States.

[5]Department of Materials Science and Engineering, Stanford University, Stanford, CA 94305, United States.

[6]Kavli Institute at Cornell for Nanoscale Science, Cornell University, Ithaca, NY 14853, United States.

*e-mail: bwang87@stanford.edu; hyhwang@stanford.edu.

## Abstract

The recent observation of superconductivity in thin film infinite-layer nickelates[1–3] offers fresh terrain for investigating superconductivity in layered oxides[4]. A wide range of perspectives[5–10], emphasizing single- or multi-orbital electronic structure, Kondo or Hund's coupling, and analogies to cuprates, have been theoretically proposed. Clearly further experimental characterization of the superconducting state is needed to develop a foundational understanding of the nickelates. Here we use magnetotransport measurements to probe the superconducting anisotropy in Nd$_{0.775}$Sr$_{0.225}$NiO$_2$. We find that the upper critical field is surprisingly isotropic at low temperatures, despite the layered crystal structure. We deduce that under magnetic field, superconductivity is strongly Pauli-limited, such that the paramagnetic effect dominates over orbital de-pairing. Underlying this isotropic response is a substantial anisotropy in the superconducting coherence length, at least four times longer in-plane than out-of-plane. A prominent low-temperature upturn in the upper critical field indicates the likelihood of an unconventional ground state.

**Main**

A key characteristic of layered superconductors is an upper critical field $H_{c2}$ often exhibiting substantial anisotropy $\gamma \equiv H_{c2,\parallel}/H_{c2,\perp}$, where $H_{c2,\parallel}$ and $H_{c2,\perp}$ indicate magnetic fields in the *a-b* plane and along the crystal *c*-direction, respectively. In layered oxides such as cuprates[11], ruthenates[12], and cobaltates[13], the anisotropy reflects the two-dimensional (2D) nature of both the crystal and electronic structure. The case for the recently discovered infinite-layer nickelates is *a priori* interesting: although it shares the same crystal structure as infinite-layer cuprates, calculations indicate both a quasi-2D hole band and three-dimensional (3D) electron bands are present in the nickelates for a broad range of electron interactions[5,6]. The relative importance of these bands is a subject of much interest with respect to the superconducting state. More generally, nickelates provide a new system within which to explore longstanding debates on the role of dimensionality for superconductivity.

With these motivations, we have studied the superconducting anisotropy through resistive measurements of $H_{c2}$ in thin-film $Nd_{0.775}Sr_{0.225}NiO_2$. This composition has the highest crystallinity in our recent investigations, as well as a superconducting transition temperature $T_c$ among the peak values observed across the superconducting dome[2]. As previously reported, our optimal samples are capped with $SrTiO_3$, and are bounded in thickness to below ~10 nm; above this we observe mixed-phase formation or decomposition. A high-angle annular dark-field scanning transmission electron microscopy (HAADF-STEM) image is shown in Fig. 1a. In good agreement with prior observations, *c*-axis oriented crystalline coherence is observed throughout the film, together with extended defects primarily based on vertical Ruddlesden-Popper (RP) type faults. Figure 1b, c shows the temperature-dependent resistivity $\rho(T)$ in the low temperature region, with varying magnetic field applied along the *c*-axis ($H_\perp$) and the *a-b* plane ($H_\parallel$, transverse geometry). While minimal magnetoresistance is observed in the normal state, the superconducting transition is suppressed with increasing magnetic field, allowing us to extract the temperature dependence of $H_{c2}$.

Figure 2 shows the main result of this study: the magnetic field-temperature phase diagram of $H_{c2,\perp}$ and $H_{c2,\parallel}$, down to 0.5 K. Here $T_c$ and $H_{c2}$ are identified as the resistive transition reaching 50% of the normal state, as discussed below. Overall, the most striking feature is that $H_{c2}$ is largely isotropic at low temperatures (as emphasized in Fig. 2a inset), despite the layered crystal structure, the quasi-2D nature of the $d_{x^2-y^2}$ hole band[5,6] and the thin film geometry. Does this reflect a rather isotropic electronic structure, and superconducting coherence length, as unusually found in (K,Ba)FeAs$_2$ (ref. [14])? Or does it have a different fundamental origin? Our aim is to address this central question. As we argue below, $H_{c2}$ is dominated by the isotropic spin response, rather than orbital effects, over much of the phase diagram. Such strongly Pauli-limited superconductivity in nickelates appears unique among the layered oxides.

We start by examining the magnetoresistive transition observed in Fig. 1b, c. The visible broadening can have contributions from static compositional inhomogeneity, superconducting fluctuations, and vortex motion in this type-II superconductor[1,9]. In terms of inhomogeneity, the

RP-type faults likely give rise to a spread in $T_c$. Regarding superconducting fluctuations, we fit the onset of superconductivity following the 2D Aslamazov-Larkin (AL) formulism[15,16], with representative fits shown in the insets of Fig. 1b, c. From this we obtain a corresponding $T_c^{AL}$, above which superconducting fluctuations significantly impact the resistive transition. We note that this regime also includes effects arising from static inhomogeneities.

As for vortex motion, we investigate its contribution in scaled Arrhenius plots, as shown in Fig. 3a, b. Between the normal state and the measurement noise floor, exponentially decreasing resistivity is seen for both orientations across several orders of magnitude. This is well described by thermally-activated vortex motion[17]: $\rho(T,H) = \rho_0(H) \exp(U_0/k_B T)$ where $\rho_0$ is the resistivity prefactor, $U_0$ is the activation energy and $k_B$ is Boltzmann's constant. Figure 3c shows $U_0$ extracted using this form. Similar to cuprates[17,18], $U_0$ for $H_\perp$ and $H_\parallel$ exhibits regions of power-law dependence on magnetic field. Note that the success of this fitting implies that vortex motion is relevant for $H_\parallel$, and the large difference in $U_0$ values between $H_\perp$ and $H_\parallel$ indicates substantial vortex-motion anisotropy. We define $T_c^{VM}$ as the temperature above which resistivity deviates from thermally-activated behavior, and attribute the resistive transition below $T_c^{VM}$ predominantly to vortex motion.

Summarizing the onset of fluctuations and vortex motion, we construct the phase diagrams in Fig. 4a, b. In addition to $H_c^{VM}(T)$ and $H_c^{AL}(T)$, we plot representative candidate proxies for $H_{c2}$: $H_c^{90\%}(T)$, $H_c^{50\%}(T)$ and $H_c^{1\%}(T)$, corresponding to the resistivity reaching 90%, 50%, and 1% of the normal state. Between $H_c^{VM}(T)$ and $H_c^{AL}(T)$ (extrapolating to 27% and 80% of the resistive transition for $H = 0$ T) there is a regime where the magnetoresistance can serve as a reliable indicator of $H_{c2}$, free from vortex creep or superconducting fluctuations which tend to under/over-estimate $H_{c2}$, as well as distort its functional form[19]. Hence, we use $H_c^{50\%}$ to measure $H_{c2}$; our conclusions are robust to variations within this regime aside from specific numerical values.

We next focus on magnetotransport near $T_c$, taking advantage of the simplification in the critical regime arising from diverging length scales. Here the central observation is a $T$-linear dependence of $H_{c2,\perp}$ and a $T^{1/2}$ dependence of $H_{c2,\parallel}$ (Fig. 4c). We first consider purely orbital depairing (superconductivity suppressed by the energy cost of the field-induced supercurrent), with the characteristic $T^{1/2}$ dependence of $H_{c2,\parallel}$ arising from 2D confinement. This is commonly observed for superconductors thinner than the Ginzburg-Landau (GL) coherence length[20,21], and is a reasonable possibility here, given the thickness limits for stabilizing $Nd_{0.775}Sr_{0.225}NiO_2$. In such geometries, $H_{c2}$ is well described by the linearized GL form[22] (generalized to an anisotropic, $c$-axis thickness-limited superconductor) as

$$H_{c2,\perp}(T) = \frac{\phi_0}{2\pi \xi_{ab}^2(0)}\left(1 - \frac{T}{T_c}\right), \tag{1}$$

$$H_{c2,\parallel}(T) = \frac{\sqrt{12}\phi_0}{2\pi \xi_{ab}(0)d}\left(1 - \frac{T}{T_c}\right)^{\frac{1}{2}}, \tag{2}$$

where $\phi_0$ is the flux quantum, $\xi_{ab}(0)$ is the in-plane zero-temperature coherence length, and $d$ is the superconducting thickness.

Using these forms, we extract $\xi_{ab}(0) = 42.6 \pm 0.1$ Å and $d = 225 \pm 7$ Å. In previous studies on other material systems, $d$ derived from such analysis was in good quantitative agreement with the physical thickness of the film[20,23], except for cases with dominant spin-orbit scattering[24,25], for which the thickness is underestimated. Here we deduce a thickness that is 2.3 – 3 times larger than the observed thickness of the film (Table S1, Supplementary Information), which we take to be an unphysical discrepancy. Furthermore, this analysis presumes confinement preventing vortex entry for $H_\parallel$, inconsistent with evident vortex motion in this orientation (Fig. 3c). Finally, 2D confinement of an orbital-limited superconductor usually leads to a dramatic enhancement of $H_{c2,\parallel}$ over $H_{c2,\perp}$ that is inversely proportional to thickness, unlike the response observed here (insets to Fig. 2a, b).

These observations strongly indicate a different origin for the behavior of $H_{c2,\parallel}$. In fact, if the superconductor is Pauli-limited, the paramagnetic effect can also give a $T^{1/2}$ dependence near $T_c$ (refs [26–29]), arising from the Zeeman energy gain of the normal state ($\sim \mu H$) competing against the superconducting condensate energy ($\sim H^2/8\pi$). In this case, fitting the $T^{1/2}$ dependence allows us to extract the electronic effective magnetic moment[28] near $T_c$ to be 2.37 ± 0.01 Bohr magnetons, indicating an enhanced magnetic susceptibility of the normal state electrons, as seen in other unconventional superconductors[30]. Additionally, as the purely orbital-limited $H_{c2,\parallel}$ necessarily exceeds the measured (Pauli-limited) $H_{c2,\parallel}$, the observation of vortex motion below $T_c^{VM}$ provides only an upper bound estimate of $\xi_c(0) \leq 10$ Å, or 3.3 NiO$_2$ layers (Supplementary Information), corresponding to $\xi_c(0)/\xi_{ab}(0) \leq 0.23$.

To quantitatively describe both orientations and lower temperatures beyond the critical regime, both orbital and paramagnetic effects clearly need to be considered, as captured in Werthamer–Helfand–Hohenberg (WHH) theory[31]. Recognizing the unusual nature of the prominent upturn at lower temperatures, we focus first on $T > 4$ K and fit the data (independently for each orientation) to the single-band WWH form, constrained to a second-order superconducting transition for all temperatures. We use single-band WHH given that no signatures of multi-band superconductivity were observed near $T_c$ (ref. [32]). The three fit parameters allowed to vary (Table S1, Supplementary Information) are the Maki parameter $\alpha$ (characterizing the ratio between the orbital-limited and Pauli-limited upper critical fields), the spin-orbit scattering parameter $\lambda_{so}$, and the effective mass $m_{eff}$. A number of consistency checks (Supplementary Information) confirm that: the spin-orbit scattering time is much longer than the elastic-transport scattering time (assumed in WHH); a reasonable electronic diffusion coefficient is deduced; $\xi_{ab}(0)$ is sensibly related to the clean-limit coherence length and the transport mean free path; and $\alpha$ is consistent with estimates from the critical regime near $T_c$.

Such overall consistency and the high quality of the fits allow us to draw some qualitative conclusions robust to the fitting details. The most prominent feature is the high values of $\alpha$ for both orientations, indicating the strong presence of the paramagnetic effect. In particular, the exceptionally large $\alpha_\parallel$ supports the purely Pauli-limited scenario for $H_{c2,\parallel}$. Such an interpretation also explains the decreasing anisotropy of $H_{c2}$, as the apparently isotropic paramagnetic effect plays an increasingly important role approaching 4 K. This transition is visible through the substantial suppression of $H_{c2,\perp}$ below the purely orbital-limited response with decreasing temperature, as illustrated via the pair-breaking picture (Supplementary Information). The large $\alpha$ values also indicate that superconductivity here is in the dirty limit, in agreement with transport estimates[2]. Finally, we note that the enhanced susceptibility independently deduced from Fig. 4c ($H_{c2,\parallel}$) is consistent with the scale for the Pauli limit relevant for the WHH fit (with enhancements associated with $\lambda_{so}$).

Below 4 K, the anomalous upturns of $H_{c2,\perp}$ and $H_{c2,\parallel}$ require considerations beyond conventional WHH theory[19,28,32,33]. Similar upturns, of much debated origin, have been observed for a number of layered superconductors[19,34]. We are not in a position to suggest a most likely scenario, but discuss a few briefly here. The behavior is tantalizingly similar to expectations for the Fulde-Ferrell-Larkin-Ovchinnikov state[33]; however, the fact that we are in the dirty limit is inconsistent with anticipated requirements. The upturn could indicate the onset of two-band superconductivity[32], but the second-band contribution would also be expected to saturate to an isotropic, Pauli-limited value. Finally, we discuss the possible onset of a first-order superconducting transition[28,31]. This tendency is expected for a strongly Pauli-limited superconductor at low temperatures, although with different curvature[35]. If we remove the second-order restriction to our WHH fit, a comparable description of the data can be obtained in the high-temperature region, although with substantially smaller $\lambda_{so}$. Therefore, $\lambda_{so}$ should be considered under-constrained by our present data, whereas the other conclusions we draw from WHH are robust to the possibility of a first-order transition, and quite reproducible (see Supplementary Information).

In summary, our studies of superconductivity in $Nd_{0.775}Sr_{0.225}NiO_2$ indicate that $H_{c2}$ is largely isotropic at low temperatures because it is strongly Pauli-limited, although $\xi_c(0)/\xi_{ab}(0) \leq 0.23$. It is thus fortuitous that we can extract quasi-intrinsic properties via measurements of thickness-limited films. The microscopic origins of the enhanced magnetic susceptibility remain to be understood and may reflect electronic correlations and proximate magnetic instabilities. It is intriguing that the low-temperature behavior of $H_{c2,\parallel}$ extrapolates to the bare Pauli limit, suggesting that the anomalous low-temperature upturn and enhanced susceptibility may have a common origin.

## Methods

### Thin film synthesis.

The samples were grown on $SrTiO_3$ substrates using pulsed-laser deposition under growth and reduction conditions previously reported[2,36].

### HAADF-STEM characterization.

Electron transparent STEM samples were prepared on a Thermo Fischer Scientific Helios G4 X focused ion beam (FIB) using the standard lift-out method, as was previously reported[2,37]. Samples were thinned to <30 nm with 2 kV Ga ions, followed by a final polish at 1 kV to reduce effects of surface damage. HAADF-STEM characterization of the sample was then obtained on an aberration-correction FEI Titan Themis at an accelerating voltage of 300 kV with a convergence angle of 30 mrad and inner collection angle of 68 mrad.

### Transport measurements.

The samples were contacted using wire-bonded aluminum wires in a 6-point Hall bar geometry. The 50% resistive criterion is applied to both temperature-sweep and field-sweep data to determine $T_c/H_{c2}$.

**Acknowledgements**

We thank R. L. Greene, A. Kapitulnik, S. A. Kivelson, P. B. Littlewood, B. Maiorov, G. A. Sawatzky, H. Takagi, R. Thomale, A. Viswanathan, and Y.-H. Zhang for discussions. The work at SLAC and Stanford was supported by the US Department of Energy, Office of Basic Energy Sciences, Division of Materials Sciences and Engineering, under contract number DE-AC02-76SF00515, and the Gordon and Betty Moore Foundation's Emergent Phenomena in Quantum Systems Initiative through grant number GBMF4415 (synthesis equipment). B.H.G. and L.F.K. acknowledge support by the Department of Defense Air Force Office of Scientific Research (No. FA 9550-16-1-0305). This work made use of the Cornell Center for Materials Research (CCMR) Shared Facilities, which are supported through the NSF MRSEC Program (No. DMR-1719875).

The FEI Titan Themis 300 was acquired through No. NSF-MRI-1429155, with additional support from Cornell University, the Weill Institute, and the Kavli Institute at Cornell. The Thermo Fisher Helios G4 X FIB was acquired with support from the National Science Foundation Platform for Accelerated Realization, Analysis, and Discovery of Interface Materials (PARADIM) under Cooperative Agreement No. DMR-1539918.

**Author Contributions**

D.L., K.L., and M.O. grew the nickelate films, and conducted the reduction experiments and structural characterization. B.H.G. and L.F.K. conducted electron microscopy. B.Y.W. performed the transport measurements, and analysis with M.R.B. B.Y.W., S.P.H., and H.Y.H. wrote the manuscript with input from all authors.

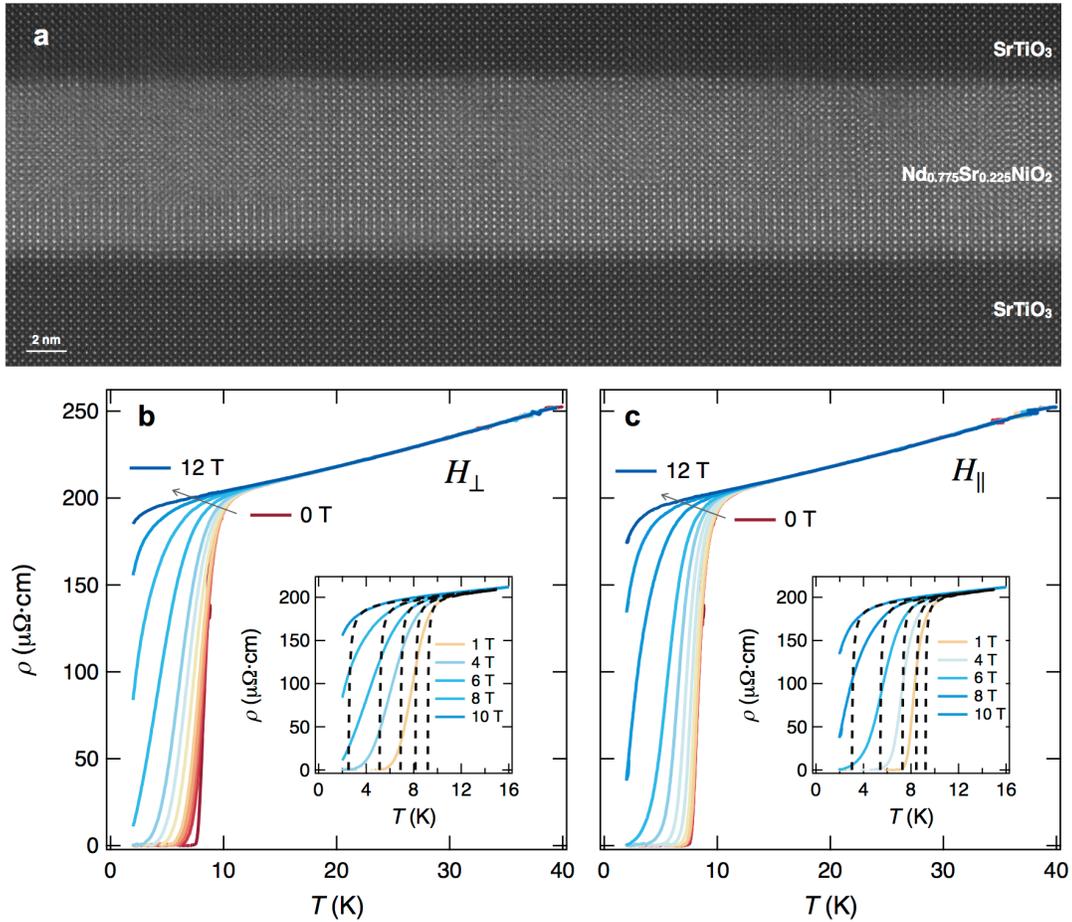

**Figure 1: Structure and magnetotransport properties of thin film Nd$_{0.775}$Sr$_{0.225}$NiO$_2$.**

**a,** Cross-sectional HAADF-STEM image of a Nd$_{0.775}$Sr$_{0.225}$NiO$_2$ thin film. **b, c,** $\rho(T)$ below 40 K at $H = 0, 0.2, 0.4, 0.6, 0.8, 1, 2, 3, 4, 6, 8, 10, 12$ T for magnetic field along the *c*-axis (**b**) and transverse in the *a-b* plane (**c**). The insets show representative resistive transitions with 2D Aslamazov-Larkin fits shown as dashed lines for field along the *c*-axis (**b**) and in the *a-b* plane (**c**).

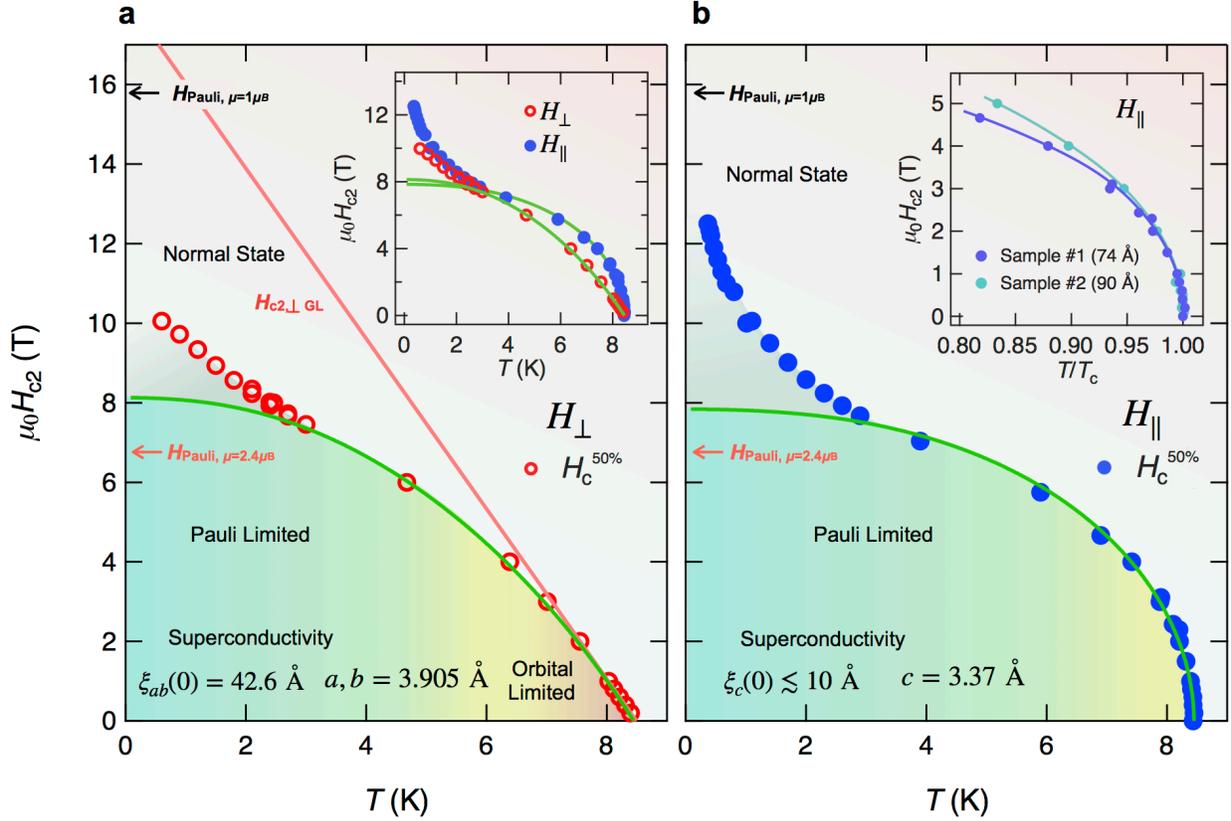

**Figure 2: $H_{c2}$ versus $T$ phase diagram for $Nd_{0.775}Sr_{0.225}NiO_2$.**

**a, b,** $H_{c2} - T$ phase diagrams for magnetic field along the $c$-axis (**a**) and in the $a$-$b$ plane (**b**). The solid green lines are Werthamer–Helfand–Hohenberg (WHH) fits of the $H_{c2}$ data above the low temperature upturn, constrained to 2$^{nd}$ order superconducting transitions at all temperatures. The solid red line (**a**) is a 3D Ginzburg-Landau fit of the $H_{c2,\perp}$ data near $T_c$. The black arrows indicate the free-electron Pauli-limited field $H_{Pauli, \mu=1\mu_B} = 1.86 \times T_c(H = 0\,T)$. The red arrows indicate the effective Pauli-limited field $H_{Pauli, \mu=2.4\mu_B}$, corresponding to the enhanced magnetic susceptibility deduced near $T_c$. The inset of (**a**) overlays the $H_{c2}$ data and the corresponding WHH fits for both field orientations. The inset of (**b**) overlays the $H_{c2,\parallel}$ data of two samples with different thickness. The solid lines are guides to the eye.

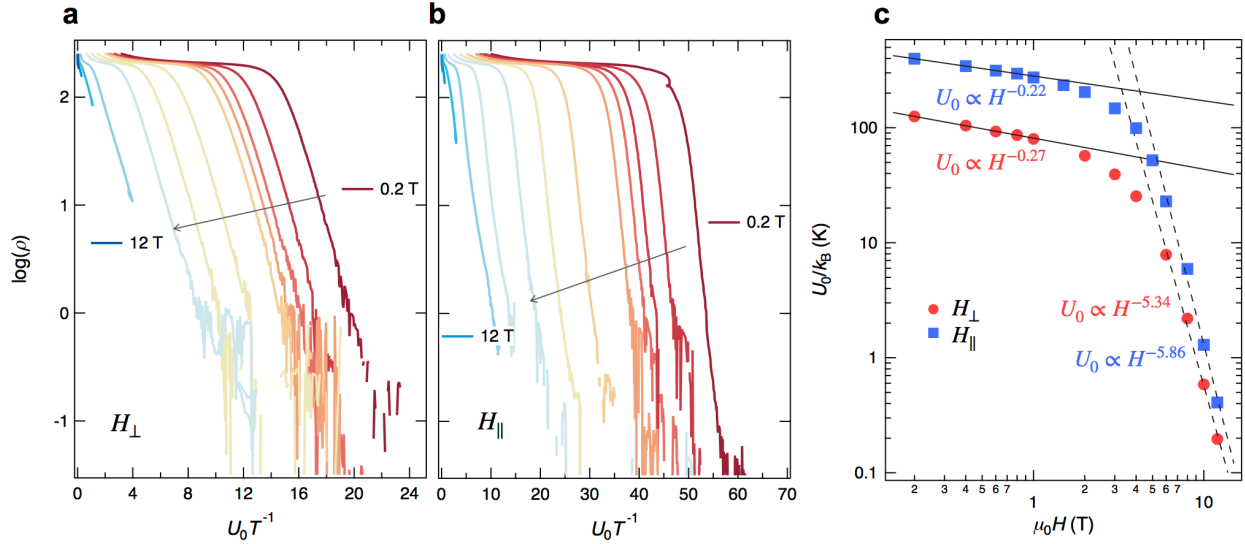

**Figure 3: Transport signatures of vortex motion, and the magnetic field dependence of the activation energy.**

**a, b,** $\rho(T)$ in scaled Arrhenius plots at field $H$ = 0.2, 0.4, 0.6, 0.8, 1, 2, 3, 4, 6, 8, 10, 12 T for magnetic field along the $c$-axis (**a**) and in the $a$-$b$ plane (**b**). The $x$-axes are scaled by the activation energies $U_0$ obtained through fits of the resistive transition in the thermally activated dissipation region according to: $\rho(T,H) = \rho_0(H) \exp(U_0/k_B T)$. **c,** The extracted $U_0$ as a function of magnetic field in double logarithmic scale for both field orientations. The solid lines are power law fits of $U_0$ below 2 K for field along the $c$-axis and in the $a$-$b$ plane, while the dashed lines are power law fits of $U_0$ above 5 K.

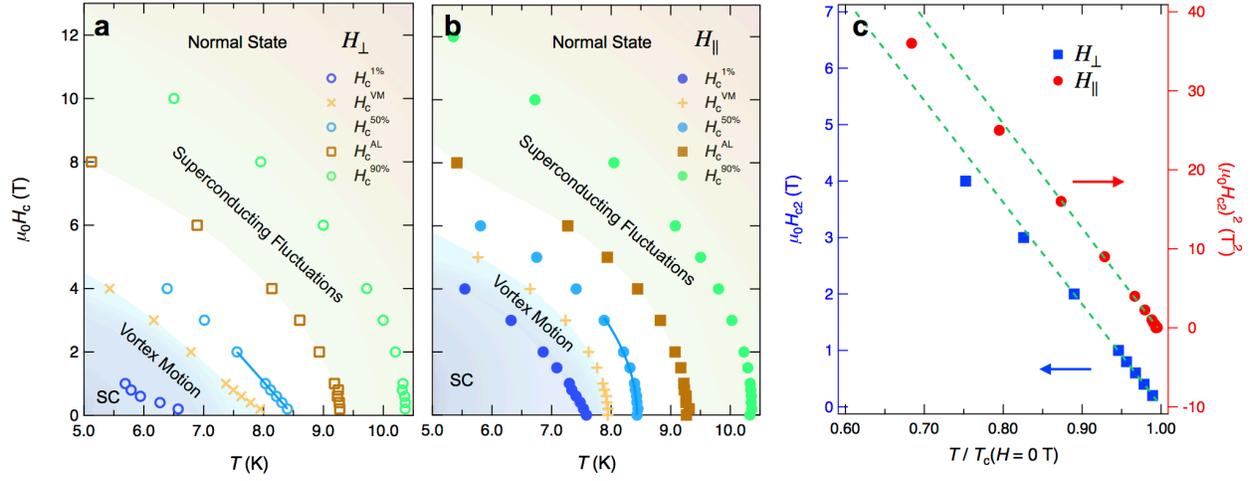

**Figure 4: $H_c$ versus $T$ phase diagram and temperature dependence of $H_{c2}$ near $T_c$.**

**a, b,** $H_c - T$ phase diagrams for magnetic field along the *c*-axis (**a**) and in the *a-b* plane (**b**). The regions separated by $H_c^{1\%}$, $H_c^{VM}$, $H_c^{AL}$, and $H_c^{90\%}$ indicate superconductivity (SC), vortex motion, superconducting fluctuations, and the normal state. **c,** Temperature dependence of $H_{c2}$ (for which we use $H_c^{50\%}$) and $(H_{c2})^2$ near $T_c$ for field along the *c*-axis and in the *a-b* plane, respectively. The dashed lines are linear fits of $H_{c2}$ and $(H_{c2})^2$ above 85% of $T/T_c(H = 0\ T)$. These fits are also shown in (**a**) and (**b**).

# Supplementary Information for
# "Isotropic Pauli-Limited Superconductivity in the Infinite Layer Nickelate Nd$_{0.775}$Sr$_{0.225}$NiO$_2$"

Bai Yang Wang,*[1,2] Danfeng Li,[2,3] Berit H. Goodge,[4] Kyuho Lee,[1,2] Motoki Osada[2,5], Shannon P. Harvey[2,3], Lena F. Kourkoutis,[4,6] Malcolm R. Beasley[3], and Harold Y. Hwang*[2,3]

[1]Department of Physics, Stanford University, Stanford, CA 94305, United States.

[2]Stanford Institute for Materials and Energy Sciences, SLAC National Accelerator Laboratory, Menlo Park, CA 94025, United States.

[3]Department of Applied Physics, Stanford University, Stanford, CA 94305, United States.

[4]School of Applied and Engineering Physics, Cornell University, Ithaca, NY 14853, United States.

[5]Department of Materials Science and Engineering, Stanford University, Stanford, CA 94305, United States.

[6]Kavli Institute at Cornell for Nanoscale Science, Cornell University, Ithaca, NY 14853, United States.

*e-mail: bwang87@stanford.edu; hyhwang@stanford.edu.

I. Reproducibility

The measurement results and conclusions drawn in the main text are reproducibly observed. We show in Fig. S1 the temperature dependent resistivity of a second sample, Sample #2, grown under the same synthesis and reduction conditions. The resistive transition shift and broadening are qualitatively identical to Sample #1 reported in the main text. Figure S2 shows the $H_{c2}$ versus $T$ phase diagrams of Sample #2 for both field orientations. Again, we observe a $T^{1/2}$ behavior of $H_{c2,\|}$ near $T_c$, a decreasing anisotropy of $H_{c2}$ at lower temperatures (shown also in the inset of Fig. S1a), and anomalous upturns of $H_{c2}$ approaching 0 K. A summary of sample and fitting parameters is given in Table S1.

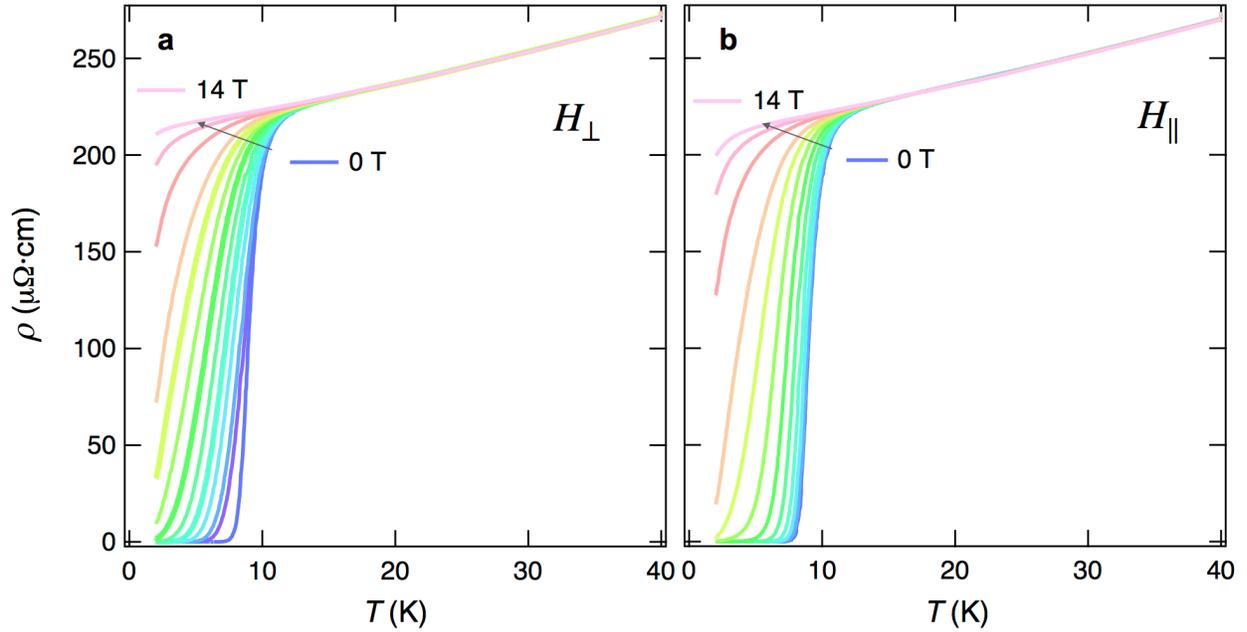

FIG. S1. $\rho(T)$ for Sample #2 below 40 K at $H$ = 0, 0.5, 1, 2, 3, 4, 5, 6, 7, 8, 10, 12, 14 T for magnetic field along the $c$-axis (**a**) and the $a$-$b$ plane (**b**).

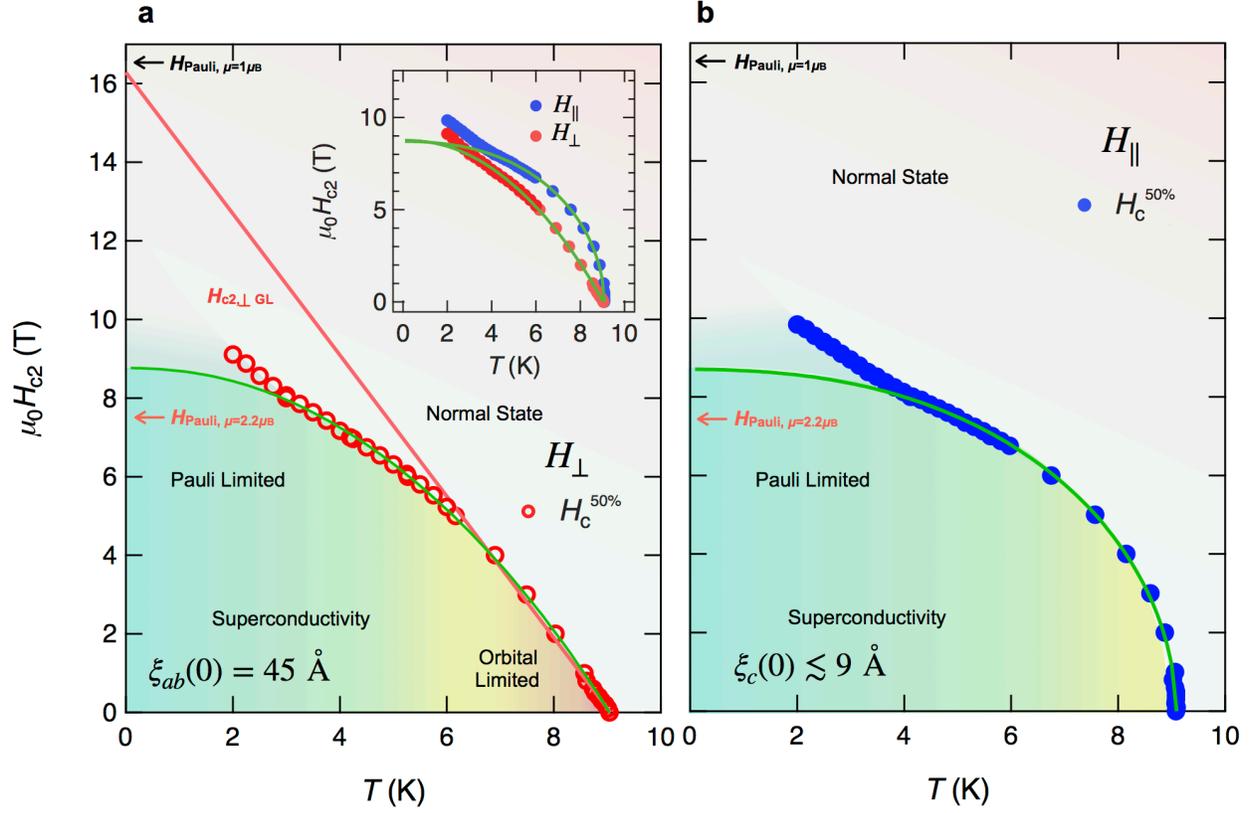

FIG. S2. **a, b,** $H_{c2} - T$ phase diagrams of Sample #2 for magnetic field along the *c*-axis (**a**) and the *a-b* plane (**b**). The solid green lines are WHH fits of the $H_{c2}$ data above the low temperature upturn. The solid red line (**a**) is a 3D Ginzburg-Landau fit of the $H_{c2,\perp}$ data near $T_c$. The black arrows correspond to the free-electron Pauli-limited field[1] $H_{\text{Pauli},\mu=1\mu_B} = 1.86 \times T_c(H=0\,\text{T})$. The red arrows correspond to the effective Pauli-limited field[2] $H_{\text{Pauli},\mu=2.2\mu_B}$, corresponding to the enhanced magnetic susceptibility deduced near $T_c$. The inset of (**a**) overlays the $H_{c2}$ data and the corresponding WHH fits for both field orientations.

## II. Estimation of the Upper Bound of $\xi_c(0)$

Considering the 3D GL anisotropic formula[1]:

$$H_{c2,\|} = \frac{\phi_0}{2\pi \xi_{ab}(0)\xi_c(0)}\left(1 - \frac{T}{T_c}\right)$$

an upper bound for $\xi_c(0)$ can be provided by the observation of vortex motion dissipation for $H_\|$. While we expect a relatively uniform superconducting order parameter for a thickness-limited ($d/1.8 < \xi_c$) 2D superconductor with no vortex motion[1], we observe a substantial region of thermally activated dissipation below $T_c^{VM}$ (Fig. S3, dashed orange line). Consequently, our observation requires $\xi_c$ to be in the 3D regime ($\xi_c < d/1.8$) below $T_c^{VM}$. This provides an upper bound of the zero-temperature c-axis coherence length: $\xi_c(0) \leq 10$ Å (9 Å for Sample #2), as shown in Fig. S3.

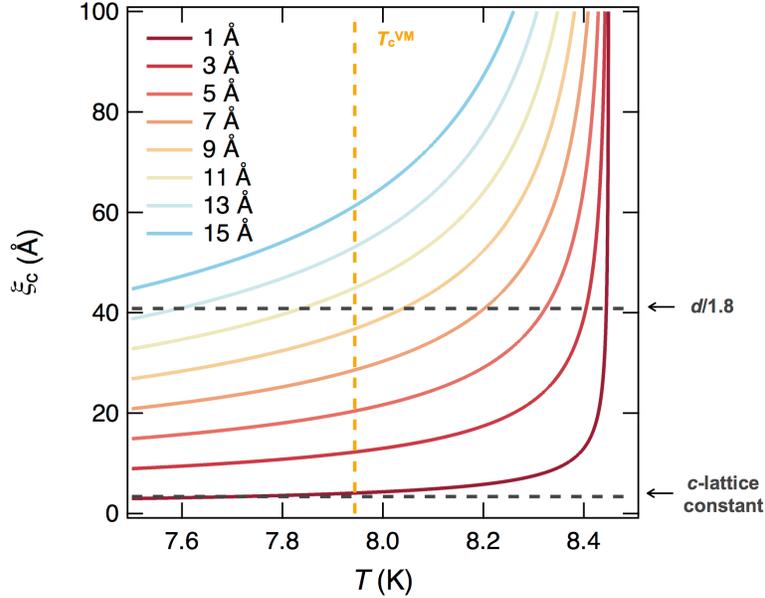

FIG. S3. Temperature dependence of the c-axis Ginzburg-Landau coherence length $\xi_c = \xi_c(0) \times (1 - T/T_c)^{-1/2}$, for $\xi_c(0)$ ranging from 1 Å to 15 Å. The vertical dashed orange line indicates $T_c^{VM}(H = 0$ T), below which thermally activated dissipation is observed for the $H_\|$ orientation. The top horizontal dashed line indicates the 2D-3D thickness crossover for vortex motion, corresponding to $d/1.8$ where $d$ is the sample thickness. The bottom horizontal dashed line indicates the c-lattice constant of the sample.

### III. Werthamer–Helfand–Hohenberg (WHH) Fitting

*WHH fit parameter self-consistency checks:* Firstly, WHH theory relies on the spin-orbit scattering time being much longer than the elastic transport scattering time. To estimate the transport scattering rate, we performed separate Hall effect measurements of the same sample in $H_\perp$ orientation. For the undoped infinite layer nickelate, the broad consensus of electronic structure calculations is that both hole and electron bands are present in the system[4–10]. As Sr doping is increasingly introduced, the electron pockets deplete. Although it is unclear when precisely the electron bands are fully depleted, we chose the highest superconducting doping level of 22.5%, where the low-temperature Hall coefficient is positive, to maximize the likelihood of a single hole-band scenario. Assuming simple Boltzmann transport and $m_{\text{eff}}$, the obtained transport scattering time is $\tau_{\text{tr}} \approx 5.9 \times 10^{-16}$ s, which is indeed much shorter than the long spin-orbit scattering time $\tau_{\text{so}} \approx 1.2 - 3.6 \times 10^{-13}$ s derived from the fitted $\lambda_{\text{so}}$ for the two field orientations, via $\lambda_{\text{so}} = h/6\pi^2 k_B T_c \tau_{\text{so}}$, where $h$ is Planck's constant. Similarly, we can estimate the normal state in-plane electronic diffusion coefficient $D_\parallel$ based on the transport scattering time and the Fermi velocity $v_f$ of the hole-band estimated from band calculations[5]: 0.6 cm$^2$/s. This result agrees well with the fitted electronic diffusion coefficient 0.4 cm$^2$/s, derived from the fitted $\alpha_\perp$, according to $\alpha = 3h/4\pi m_{\text{eff}} v_f^2 \tau_{\text{tr}}$. In addition, we note that the GL fit deduced in-plane coherence length $\xi_{ab}(0) \sim 42.6 \pm 0.1$ Å matches the geometric mean of the clean-limit coherence length $\xi_0 = 0.18 \times hv_f/2\pi k_B T_c \sim 921$ Å and the transport mean free path $l_{\text{mfp}} \sim 3.3$ Å, as expected. Finally, the Maki parameter $\alpha$ can also be estimated from linear fits of $H_{c2}$ near $T_c$ (ref. [11]): $\alpha_\parallel = 20.6$, $\alpha_\perp = 1.1$, in reasonable agreement with those obtained from the WHH fits (see Table S1). Based on the above estimates, it is worth noting that for our system, $k_f l_{\text{mfp}} \sim 3.8$ (where $k_f$ is the Fermi momentum[5]) and $l_{\text{mfp}} \ll \xi_{ab}(0)$, reiterating that here superconductivity is in the dirty limit. (A similar $k_f l_{\text{mfp}} \sim 4.2$ results assuming a 2D free electron gas.) This raises the possibility of localization effects on superconductivity[12,13], with regards to the anomalous upturn of $H_{c2}$. However, it is unclear how localization effects are manifested in a strongly Pauli-limited superconductor.

*2nd order versus 1st order WHH fits:* Due to the unconventional nature of the anomalous upturns of $H_{c2}$ at low temperatures[14], we limit our conventional WHH-theory-based fits to the high temperature regime. However, this leaves ambiguous, and experimentally under-constrained, the detailed WHH form at low temperatures. Therefore, we present two extremal cases here: 1) constraining the WHH fit to 2nd order superconducting transitions at all temperatures (as in main text); and 2) allowing for 1st order superconducting transitions to occur at low temperatures. The obtained WHH fits are shown in Fig. S4 as solid green and orange lines and the corresponding fit parameters are shown in Table S1. Although there is substantial variation in the fitted $\lambda_{so}$, which largely determines the order of the low-temperature phase transition, the magnitude of the obtained $\alpha$ and $m_{eff}$ are fairly independent of the fit choices, allowing robust qualitative conclusions to be drawn from these parameters. Note that the WHH form assumes a fixed *g*-factor of 2, relevant for comparison to the susceptibility enhancement deduced near $T_c$.

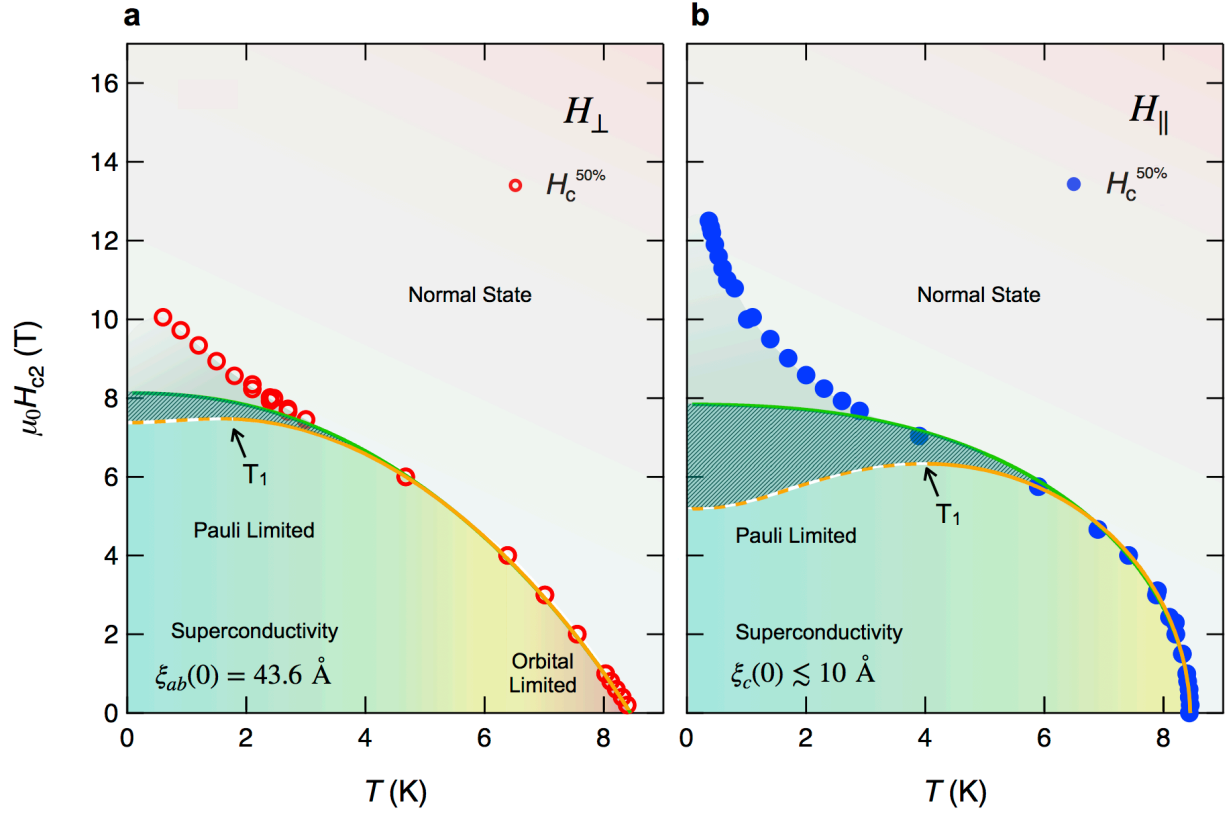

FIG. S4. **a, b,** $H_{c2} - T$ phase diagrams of Sample #1 (main text) for magnetic field along the *c*-axis (**a**) and the *a-b* plane (**b**). The solid green lines are WHH fits of the $H_{c2}$ data above the low temperature upturn, constrained to 2$^{nd}$ order superconducting transitions at all temperatures, and the orange lines are WHH fits allowing 1$^{st}$ order transitions to occur at low temperatures. The dashed section of the orange line below the tricritical temperature $T_1$ corresponds to the supercooling field where the superconducting transition becomes 1$^{st}$ order[2,11,15].

IV. Paramagnetic versus Orbital Effects for $H_\perp$ Orientation

The relative contributions from paramagnetic effects and orbital effects to the temperature dependence of the upper critical field can be identified using the pair-breaking form[1]: $\ln(T/T_c) = \psi(1/2) - \psi(1/2 + \beta/2\pi k_B T_c)$, where $\psi$ is the digamma function and $\beta$ is the pair-breaking process energy. The $\beta$ values corresponding to paramagnetic and orbital effects with perpendicular magnetic field are $\tau_{so} e^2 h H^2/4\pi m_0^2$ and $D_\parallel e H/c$, respectively. Here $e$ is the electronic charge, $c$ is the speed of light, and $m_0$ is the free electron mass. The $\beta$ values of the two effects can be added independently when considered together[1].

As shown in Fig. S5, the pair-breaking formula considering only orbital effects (solid red line) fits $H_{c2,\perp}$ above 7 K satisfactorily. The extracted electronic diffusion coefficient $D_\parallel \approx 0.5$ cm$^2$/s matches that obtained from both the WHH fit and the transport measurements discussed above. Below 7 K, however, $H_{c2,\perp}$ clearly deviates below the pair-breaking formula fit, suggesting an increasing role of the paramagnetic effect. Shown as a solid green line in Fig. S5, the pair-breaking formula fit considering both paramagnetic and orbital effects describes the temperature dependence of $H_{c2,\perp}$ accurately down to the lower temperature upturn. This supports the identification that $H_{c2,\perp}$ is orbital-limited near $T_c$ but crosses over to become Pauli-limited at lower temperatures.

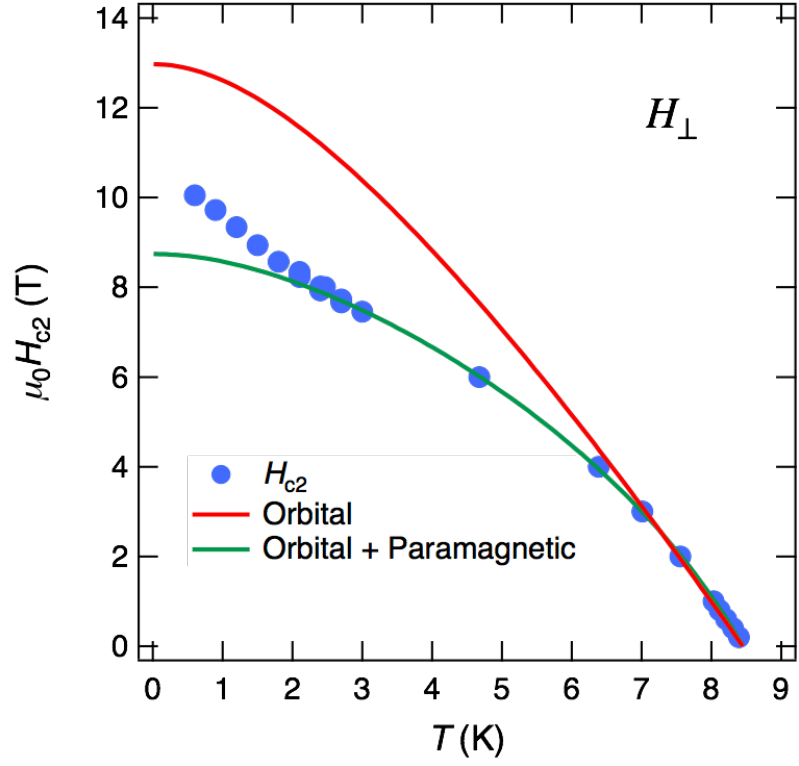

FIG. S5. Upper critical field $H_{c2}$ versus $T$ phase diagram of Sample #1 (main text) for $H_\perp$ orientation. Blue dots are measured $H_{c2,\perp}$. Solid red and green lines are pair-breaking formula fits considering only orbital effects, and considering both orbital and paramagnetic effects, respectively.

V. TABLE S1. Parameters of the sample, superconducting phase, and WHH fits for Sample #1 and #2. Note that the form of the WHH fit becomes insensitive to the precise values of $\alpha$ as the superconductivity becomes strongly Pauli-limited ($\alpha \gg 1$). We further note that $m_{eff}$ here is an effective parameter following the starting assumptions of WHH – i.e. of BCS superconductivity in a Fermi liquid with a spherical Fermi surface, thus not directly comparable to the band-structure or quasiparticle effective mass.

| | $T_c$ (K) | Sample thickness (Å) | Extracted thickness $d$ (Eq. 2) (Å) | Extracted $\xi_{ab}(0)$ (Å) | Effective moment ($\mu_B$) | Maki parameter $\alpha$ (2nd order/ 1st order) | Spin-orbit scattering parameter $\lambda_{so}$ (2nd order/ 1st order) | Effective mass $m_{eff}$ (2nd order/ 1st order) |
|---|---|---|---|---|---|---|---|---|
| Sample #1 $H_\perp$ | 8.5 | 74 | 225 | 43 | 2.4 | 1.87/1.6 | 0.27/0 | 0.67/0.78 |
| Sample #1 $H_\parallel$ | | | | | | 235.5/78.8 | 0.82/0 | 0.33/0.46 |
| Sample #2 $H_\perp$ | 9.0 | 90 | 205 | 45 | 2.2 | 1.59/1.3 | 0.26/0 | 0.73/0.87 |
| Sample #2 $H_\parallel$ | | | | | | 97.3/59.5 | 0.88/0.14 | 0.33/0.44 |